\documentclass[12pt]{article}%,leqno,report
\usepackage[cp1251]{inputenc}  % WINDOWS encoding
\usepackage[english]{babel}
\usepackage{amsmath,amssymb,epsfig,graphicx,doc,latexsym,amsfonts,bm,longtable}% ,mychap
\usepackage[bookmarksopen,colorlinks]{hyperref}
\usepackage[monochrome]{color}
 \hypersetup{colorlinks=true}

\begin{document}
\begin{center}
{\textbf{\LARGE{Renormalized theory of the ion cyclotron turbulence
in magnetic field--aligned  plasma shear flow}}}

\bigskip
\bigskip
{\large{V.S. Mikhailenko$^{\dag,\ddag}$}}\footnote{Electronic mail:
\href{mailto:vmikhailenko@kipt.kharkov.ua}{vmikhailenko@kipt.kharkov.ua}},
{\large{V.V. Mikhailenko$^{\dag,\ddag}$, K.N.
Stepanov$^{\dag,\ddag}$} and {\large{N.A. Azarenkov$^{\dag}$}}}
\\
\bigskip $^{\dag}${\textit{Kharkov National University, 61108 Kharkov,
Ukraine}}

$^{\ddag}${\textit{National Science Center ``Kharkov Institute of
Physics and Technology'',
\\
61108 Kharkov, Ukraine}}

\end{center}

\begin{abstract}
The analytical treatment of nonlinear evolution of the
shear-flow-modified current driven ion cyclotron instability and
shear-flow-driven ion cyclotron kinetic instabilities of magnetic
field--aligned plasma shear flow is presented. Analysis is performed
on the base of the nonlinear dispersion equation, which accounts for
a new combined effect of plasma turbulence and shear flow. It
consists in turbulent scattering of ions across the shear flow with
their convection by shear flow and results in enhanced nonlinear
broadening of ion cyclotron resonances. This effect is found to lead
to the saturation of ion cyclotron instabilities as well as to the
development of nonlinear shear flow driven ion cyclotron
instability.
\\ 52.35.Ra
\end{abstract}

\newpage

\section*{I. INTRODUCTION.}

The important feature of the near-Earth space plasma is its
irregularities \cite{Fejer-1980, McFadden} in density, temperature
of ions and electrons, currents and electric field. The existence of
inhomogeneous flows and currents along and across Earth's magnetic
field is a fundamental characteristic of space plasmas. One
important element of the ionospheric flows appears to be a velocity
shear. Compared to the homogeneous plasma case, sheared flows along
the magnetic field introduce significant modifications to plasma
stability. The presence of magnetic field--aligned plasma shear
flows give rise to numerous instabilities of plasmas, which are
predicted theoretically and observed experimentally in a broad
frequency range with applications to ionospheric and fusion plasmas.
The interest in scrutinizing the shear flow driven instabilities,
that are still under intense investigation (see, e.g., recent papers
\cite{Mikhailenko2006}-\cite{Mikhailenko2007} and references
therein), lies, in part, in a number of observations which indicate
correlation in place and time of broadband, low--frequency waves,
and transverse ion energization with sheared flows on the boundaries
of plasma structures\cite{Amatucci-1999}. Plasma shear flows were
often encountered by Prognoz-8 satellite \cite{Bleski} near the high
latitude magnetopause which is the boundary between the Earth's
magnetic field and solar wind. It was found\cite{Bleski} that such
plasma flows are always associated with strong wave activity at ion
cyclotron and lower hybrid frequencies ranges. High time resolution
measurements of ion distribution by Fast Auroral Snapshot (FAST)
satellite have revealed\cite{McFadden, Carlson} at an altitude of
4000km narrow ion and electron beams with steep transverse spatial
gradients in their energy. These beams were clearly identified as
the sources of free energy to drive observed electrostatic and
electromagnetic ion cyclotron waves\cite{McFadden1, Cattell} and
particles energization in regions containing ion beams and
field-aligned currents. Electrostatic IC waves have been invoked in
the explanations of the transverse anomalously strong heating of
ions in ionosphere, that cannot be accounted for by frictional
(Joule) heating \cite{Okuda, Kintner}.

The field-aligned currents\cite{Kindel}, ion
beams\cite{Bergmann-1984} or relative streaming between ion species
\cite{Bergmann-1988}, have been proposed to provide the free energy
for the development of IC instabilities. There are some \textit{in
situ} ionospheric observations \cite{Bering, Kintner-1989} that
support the development of the classical current driven
electrostatic IC instability \cite{Drummond-1962} in ionosphere.
(Comprehensive review of the theoretical, numerical and laboratory
experimental investigations of the current driven electrostatic IC
instability was given in Ref. \cite{Rasmussen}). However, generally
the levels of field--aligned current at which IC waves and
transverse ion heating were observed in ionosphere were subcritical
for the development of this instability, which has the lowest
threshold for current density in the ionospheric plasma environment
\cite{Kindel}. Kinetic effects, such as finite ion Larmor radius
effects,  electron Landau damping and ion cyclotron damping, are
pronounced for ion cyclotron modes and kinetic theory is requisite
for treating these modes. In Ref \cite{Hirose} the linear dispersion
equation for plasma with the magnetic field aligned flows with
velocities ${\bf{V}}_{0\alpha}\left(x\right)\|
{\bf{B}}\|{\bf{e}}_{z}$ of plasma components of $\alpha$ species
($\alpha=i$ for ions and $\alpha=e$ for electrons) with
inhomogeneous number densities $n_{0\alpha}\left(x \right)$ was
obtained using a kinetic approach in the local approximation, for
which $k_x L_{n} \gg 1$ and $k_x L_v \gg 1$, where $L_n = \left[
d\ln n_0 \left(x \right)/dx \right]^{-1}$,  $L_v= \left[ d\ln V_0
\left(x\right)/dx \right]^{-1}$, and $k_x$ is the projection of the
wave vector $\bf{k}$ normal to the flow velocity and along the
velocity gradient. That equation may be presented in a form (see,
also papers
\cite{Mikhailenko2006},\cite{Hirose}-\cite{Ganguli-2002}) as
\begin{equation}
\label{1}\varepsilon \left( {\bf{k},\omega}\right)= 1 +
\sum\limits_{\alpha = i,e} {\delta \varepsilon _\alpha}\left(
{\bf{k},\omega } \right) = 1 + \sum\limits_{\alpha = i,e}{\left(
{\delta \varepsilon _\alpha ^{\left( 1 \right)} \left(
{\bf{k},\omega } \right) + \delta \varepsilon _\alpha ^{\left(2
\right)} \left( {\bf{k},\omega } \right)} \right)} = 0.
\end{equation}
In Eq.(\ref{1}) $\delta\varepsilon_{\alpha}^{\left( 1 \right)}
\left(\bf{k},\omega\right)$ is the conventional dielectric
permittivity of the $\alpha$-species of the plasma without shear
flow and $\delta \varepsilon _\alpha ^{\left( 2 \right)} \left(
\bf{k},\omega\right)$ is the velocity shear dependent part of the
dielectric permittivity $\delta \varepsilon _\alpha \left(
\bf{k},\omega\right)$. Here and in what follows
${\bf{k}}=\left(k_{x}, k_{y}, k_{z}\right)=\left(k_{\bot}, \theta,
k_{z}\right)$, where $k_{x}=k_{\bot}\cos \theta$,
$k_{y}=k_{\bot}\sin\theta$, and $k_{z}$ is the projection of the
wave vector on the magnetic field ${\bf{B}}$. It was assumed in
\cite{Mikhailenko2006},\cite{Hirose}-\cite{Ganguli-2002} the
equilibrium distribution function $F_{\alpha 0}$ to be a drifting
Maxwellian (see below Eq.(\ref{26})), even though this may be
somewhat unrealistic for the collisionless ionosphere. The
Maxwellian assumption permits calculationally convenient and
comprehensive exploration of the IC instabilities. Eq.(\ref{1}) was
investigated for high frequency instabilities\cite{Hirose} with
frequencies $\omega$ well above the IC frequency $\omega_{ci}$, as
well as for low frequency instabilities\cite{Gary, Lakhina} with
frequencies below $\omega_{ci}$. The investigation of the IC
instabilities in magnetic field--aligned  plasma shear flows, that
is a focus of this paper, was initiated in Ref.\cite{Lakhina}, where
IC instability driven by velocity shear of hot ion beam in plasma
with cold ions and electrons was considered. Thereafter it was
derived in Ref.\cite{Belova} that plasma shear flow along the
magnetic field may be unstable against the development of an IC
instability of the hydrodynamic type with wavelength much less than
ion thermal Larmor radius $\rho_{i}= v_{Ti }/\omega _{ci}$. That
instability was developed even when the velocity of the electron-ion
drift along the magnetic field was below the threshold for current
driven IC instability.

The comprehensive investigation of the IC instabilities of
magnetic-field aligned plasma shear flow was undertaken in
Ref.\cite{Mikhailenko2006}. It was shown analytically that under
conditions
\begin{eqnarray}
\displaystyle|\text{Re}\,\varepsilon_{i}^{\left(1\right)}\left(\mathbf{k},
\omega\right)|\sim
|\text{Re}\,\varepsilon_{i}^{\left(2\right)}\left(\mathbf{k},
\omega\right)|>
|\text{Im}\,\varepsilon_{i}^{\left(2\right)}\left(\mathbf{k},
\omega\right)|\gtrsim|\text{Im}\,\varepsilon_{i}^{\left(1\right)}
\left(\mathbf{k}, \omega\right)|\label{2}
\end{eqnarray}
shear flow along the magnetic field does not only modify the
frequency, growth rate and the threshold of the known current driven
IC  instability, but it is a source of the development of the
kinetic and hydrodynamic shear-flow-driven IC instabilities at the
levels of field-aligned current which are subcritical for the
development of the current driven IC instability. Shear flow along
the magnetic field leads to the splitting of the separate IC mode,
existing in the plasma without shear flow\cite{Drummond-1962}, into
two IC modes with frequencies $\omega_{1,2}\left(\mathbf{k}\right)$,
which correspond to different kinds of the IC instabilities. In the
limiting case of IC waves, which propagate almost across the
magnetic field, the frequencies
$\omega_{1,2}\left(\mathbf{k}\right)$, which are the solutions of
the linear dispersion equation
$1+\varepsilon_{i}^{\left(1\right)}\left(
\bf{k},\omega\right)+\varepsilon_{i}^{\left(2\right)}\left(
\bf{k},\omega\right)+\varepsilon_{e} \left( \bf{k},\omega\right)=0,$
may be obtained in the form $\omega_{1,2}\left(\mathbf{k}\right)=
k_{z}V_{0i}+n\omega_{ci}+\delta\omega_{1,2}\left(\mathbf{k}\right)$,
with $|\delta\omega_{1,2}\left(\mathbf{k}\right)| \ll n\omega_{ci}$,
$|z_{i1,2}|=|\delta\omega_{1,2}\left(\mathbf{k}\right)|/\sqrt{2}|k_{z}|v_{Ti}\gg
1$,
$|z_{e}|=|n\omega_{ci}-k_{z}\left(V_{e0}-V_{i0}\right)|/\sqrt{2}|k_{z}|v_{Te}\lesssim
1$. In the limiting case of weak flow shear, for which
\begin{eqnarray}&\displaystyle
\left(n\omega_{ci}-k_{y}v_{di}\right)^{2}A_{in}\left(k_{\bot} ^{2}
\rho _{i}^{2 } \right)>4\frac{k_{y}V'_{0i}}
{k_{z}\omega_{ci}}k^{2}_{z}v_{Ti}^{2}\left(1-A_{i0}\left(k_{\bot}
^{2} \rho _{i}^{2 }\right)+\tau\right),\label{3}
\end{eqnarray}
two kinetic IC instabilities are excited. Here and in what follows
$\tau=T_{i}/T_{e}$ with $T_{e,i}$ being the electron and ion
temperatures, respectively, $\lambda_{D\alpha}^{2}$ is the Debye
radius, $A_{\alpha n}\left(k_{\bot}^{2} \rho _{\alpha}^{2} \right) =
I_{n}\left(k_ {\bot} ^{2} \rho _{\alpha} ^{2 }\right)e^{ -
k_{\bot}^{2} \rho _{\alpha} ^{2}}$, and
$v_{di}=\left(cT_{i}/eB_{0}\right)d\ln n_{i0}\left(x\right)/d x$.
The frequency
$\text{Re}\,\delta\omega_{1}\left(\mathbf{k}\right)=\delta\omega_{01}\left(\mathbf{k}\right)$,
and the growth rate,
$\gamma_{01}=\text{Im}\,\delta\omega_{1}\left(\mathbf{k}\right)=\gamma_{i1}+\gamma_{e1}$
of the first instability are equal approximately to
\begin{eqnarray}&\displaystyle
\text{Re}\,\delta\omega_{1}\left(\mathbf{k}\right)=\delta\omega_{01}\left(\mathbf{k}\right)\approx
\frac{\left(n\omega_{ci}-k_{y}v_{di}\right)A_{in}\left(k_{\bot} ^{2}
\rho _{i}^{2 } \right)} {\left(1-A_{i0}\left(k_{\bot} ^{2} \rho
_{i}^{2 } \right)+\tau\right)}, \label{4}
\end{eqnarray}
\begin{eqnarray}&\displaystyle
\gamma_{i1}=-\frac{\left(\delta
\omega_{01}\left(\mathbf{k}\right)\right)^{2}}{\Omega_{n}}k^{2}\lambda^{2}_{Di}
\text{Im}\,\delta\varepsilon_{i}\left(\mathbf{k},
n\omega_{ci}\right)\nonumber
\\ &\displaystyle=-\frac{\left(\delta
\omega_{01}\left(\mathbf{k}\right)\right)^{2}}{\Omega_{n}}A_{in}\left(k_{\bot}
^{2} \rho _{i}^{2 } \right)
\sqrt{\frac{\pi}{2}}\frac{e^{-z^{2}_{i1}}}{|k_{z}|v_{Ti}}\left[\left(n\omega_{ci}-k_{y}v_{di}\right)
-\frac{k_{y}V'_{0i}}{k_{z}\omega_{ci}} \delta
\omega_{01}\left(\mathbf{k}\right)\right],\label{5}
\end{eqnarray}
\begin{eqnarray}&\displaystyle
\gamma_{e1}=-\frac{\left(\delta
\omega_{01}\left(\mathbf{k}\right)\right)^{2}}{\Omega_{n}}\tau\sqrt{\frac{\pi}{2}}\frac{\left(n\omega_{ci}
-k_{z}\left(V_{0e}-V_{0i}\right)-k_{y}v_{de}\right)}
{|k_{z}|v_{Te}}\exp\left(-z_{e}^{2}\right),\label{6}
\end{eqnarray}
where
\begin{eqnarray}&\displaystyle
\Omega^{2}_{n}=\left(n\omega_{ci}-k_{y}v_{de}\right)^{2}A^{2}_{in}\left(k_{\bot}
^{2} \rho^{2}_{i}\right)
-4\frac{k_{y}V'_{0i}}{k_{z}\omega_{ci}}k^{2}_{z}v_{Ti}^{2}A_{in}
\left(k_{\bot} ^{2} \rho^{2}_{i} \right)\left(1-A_{i0}\left(k_{\bot}
^{2} \rho _{i}^{2 } \right)+\tau\right).\label{7}
\end{eqnarray}
It follows from Eqs. (\ref{4})-(\ref{6}), that
$\delta\omega_{1}\left(\mathbf{k}\right)$, defines the frequency and
the growth rate of the current driven IC
instability\cite{Drummond-1962}, modified by flow
shear\cite{Mikhailenko2006}. This instability develops due to
inverse electron Landau damping, when
$k_{z}\left(V_{0e}-V_{0i}\right)+k_{y}v_{de}>n\omega_{ci}$ under
condition $|z_{i0}|\gg 1$ of negligible IC damping of the mode
$\omega_{1}$.

The solution $\delta\omega_{2}\left(\mathbf{k}\right)$ defines a new
shear flow driven branch of the IC waves, which is absent in
shearless plasma flows. The frequency
$\text{Re}\,\delta\omega_{2}\left(\mathbf{k}\right)$ and growth rate
$\gamma_{02}=\text{Im}\,\delta\omega_{2}\left(\mathbf{k}\right)=\gamma_{i2}+\gamma_{e2}$
of this second instability are equal to\cite{Mikhailenko2006}
\begin{eqnarray}&\displaystyle
\text{Re}\,\delta\omega_{2}\left(\mathbf{k}\right)=
\delta\omega_{02}\left(\mathbf{k}\right)\approx\frac{k_{y}V'_{0i}}{k_{z}\omega_{ci}}\frac
{k^{2}_{z}v_{Ti}^{2}}{\left(n\omega_{ci}-
k_{y}v_{di}\right)},\label{8}
\end{eqnarray}
\begin{eqnarray}&\displaystyle
\gamma_{i2}=\frac{\left(\delta
\omega_{02}\left(\mathbf{k}\right)\right)^{2}}{\Omega_{n}}k^{2}\lambda^{2}_{Di}\text{Im}\,
\delta\varepsilon_{i}\left(\mathbf{k},
n\omega_{ci}\right)=\left(\frac{k_{y}V'_{0i}}{k_{z}\omega_{ci}}
\frac{k^{2}_{z}v^{2}_{Ti}}
{\left(n\omega_{ci}-k_{y}v_{di}\right)}\right)^{2}\nonumber
\\ &\displaystyle\times\sqrt{\frac{\pi}{2}}\frac{A_{in}\left(k_{\bot} ^{2} \rho _{i}^{2
} \right)}
{\Omega_{n}}\frac{e^{-z^{2}_{i2}}}{|k_{z}|v_{Ti}}\left[\left(n\omega_{ci}-k_{y}v_{di}\right)-
\left(\frac{k_{y}V'_{0i}}{k_{z}\omega_{ci}}
\right)^{2}\frac{k^{2}_{z}v^{2}_{Ti}}
{\left(n\omega_{ci}-k_{y}v_{di}\right)}\right],\label{9}
\end{eqnarray}
\begin{eqnarray}&\displaystyle
\gamma_{e2}=\frac{\left(\delta
\omega_{02}\left(\mathbf{k}\right)\right)^{2}}{\Omega_{n}}\tau\sqrt{\frac{\pi}{2}}\frac{\left(n\omega_{ci}
-k_{z}\left(V_{0e}-V_{0i}\right)-k_{y}v_{de}\right)}
{|k_{z}|v_{Te}}\exp\left(-z_{e}^{2}\right).\label{10}
\end{eqnarray}
As it follows from Eq.(\ref{10}), this mode becomes unstable for any
values of $k_{\bot}\rho _{i}$ due to inverse electron Landau
damping, when the velocity of the relative drift between ions and
electrons is \textit{below} the critical value
$V_{0e}^{\left(c\right)}$ \cite{Mikhailenko2006}, roughly estimated
as $V_{0e}^{\left(c\right)}= V_{0i}+
\left(n\omega_{ci}-k_{y}v_{de}\right)/k_{z}$, i.e. under conditions
at which the current driven IC instability modified by shear flow
does not develop.

In the case of a strong flow shear, for which the condition opposite
to the above presented condition (\ref{3}) is met, but with
$k_{y}V'_{0i}/ k_{z}\omega_{ci}<0$, the growth rates $\gamma_{1}$
and $\gamma_{2}$ are determined by Eqs.(\ref{5}), (\ref{6}) and
(\ref{9}), (\ref{10})), but with frequencies $\delta\omega_{01,02}$
replaced with
\begin{eqnarray}&\displaystyle
\delta\omega_{\left(+,-\right)}\approx\pm
k_{z}v_{Ti}\left(\frac{k_{y}V'_{0i}}{k_{z}\omega_{ci}}
\frac{A_{in}\left(k_{\bot} ^{2} \rho _{i}^{2 } \right)}
{\left(1-A_{i0}\left(k_{\bot} ^{2} \rho _{i}^{2 }
\right)+\tau\right)}\right)^{1/2}.\label{11}
\end{eqnarray}

In the case of a sufficiently strong flow shear, when
$\Omega^{2}_{n}<0$, the shear flow driven IC instability of the
hydrodynamic (reactive) type is excited with frequency
$\delta\omega_{\left(H\right)}$ and growth rate
$\gamma_{\left(H\right)}$ approximately equal to
\begin{eqnarray}
&\displaystyle
\delta\omega_{\left(H\right)}\simeq\frac{1}{2}\frac{\left(n\omega_{ci}-k_{y}v_{di}\right)
A_{in}\left(k^{2}_{\perp}\rho^{2}_{i}\right)}{\left(1-A_{i0}\left(k^{2}_{\perp}\rho^{2}_{i}\right)
+\tau\right)}\,, \qquad\label{12}
\end{eqnarray}
\begin{eqnarray}
&\displaystyle \gamma_{\left(H\right)}\simeq\frac{
\left(\dfrac{k_{y}}{k_{z}}\dfrac{V'_{0i}}{\omega_{ci}}k^{2}_{z}v^{2}_{Ti}
A_{in}\left(k_{\perp}^{2}\rho_{i}^{2}\right)\left(1+\tau\right)
-\frac{1}{4}\left(n\omega_{ci}-k_{y}v_{di}\right)^{2}A_{in}^{2}\left(k_{\perp}^{2}\rho
_{i}^{2}\right)\right)^{1/2}}{\left(1-A_{i0}\left(k^{2}_{\perp}\rho^{2}_{i}\right)+\tau
\right)}\,.\label{13}
\end{eqnarray}
This instability is in fact the extension of the hydrodynamic
D'Angelo instability \cite{D'Angelo} onto the IC frequency range.

An important, but still absent, element in the studies of IC
instabilities of plasmas with  parallel shear flow is an
understanding of the processes of their nonlinear evolution and
saturation. For the current driven IC instabilities of plasma
without flow shear these studies were grounded on the nonlinear
theory of the IC resonances broadening, which was developed in Ref.
\cite{Dum}. The saturation level of the current driven IC
instability obtained in Refs. \cite{Dum, Benford} appears to agree
with IC heating experiments \cite{Dakin, Rasmussen}. The presence of
shear flows leads to more complicate picture of ions scattering in
turbulent electric fields and resonance broadening saturation
mechanism. The present work extends the earlier studies of the
renormalized theory of the IC turbulence by including combined
effect of plasma turbulence and shear flow, which consists in
turbulent scattering of ions across the shear flow into the regions
with a greater or smaller flow velocity and enhancing by this means
transport of ions along shear flow. In Section II of this paper we
derive the nonlinear dispersion equation, which accounts for the
effect of IC resonances broadening resulted from the random motion
of ions in the electric field of the IC turbulence in magnetic
field--aligned plasma shear flow. In Section III we present the
approximate qualitative analysis of the nonlinear evolution of the
shear flow modified current driven IC instability and shear flow
driven kinetic IC instabilities, which resulted from the IC
resonance broadening effect in the presence of shear flow. Finally,
in Section IV we summarize our results.

\section*{II. NONLINEAR DISPERSION EQUATION}

Our theory is based on the Vlasov-Poisson system of equations. We
use leading center coordinates for ions
$X=x+\left(v_{\bot}/\omega_{ci}\right)\sin\phi$,
$Y=y-\left(v_{\bot}/\omega_{ci}\right)\cos\phi$, where $x, y, z$ are
usual local particle coordinates with $z$-axis directed along the
magnetic field $\mathbf{B}$, and where $v_{\bot}$ is velocity and
$\phi$ is gyrophase angle of the gyromotion of ion. We find it
suitable to use instead of $z$ and $\phi$ new variables
$z_{1}=z-\int\limits^{t}v_{z}\left( \tau\right)d\tau,
\phi_{1}=\phi+\omega_{ci}t$.  With these variables the governing
Vlasov equation describing the perturbation $f_{i}$ by the
self-consistent electrostatic potential $\Phi$ of the ion
distribution function $F_{i}$, $F_{i}=F_{i 0}+f_{i}$, where $F_{i
0}$ is the equilibrium function of the distribution of ions, takes
the form
\begin{eqnarray}
& \displaystyle \frac{\partial f_{i}}{\partial
t}+\frac{e}{m_{i}\omega_{ci}} \left(\frac{\partial\Phi}{\partial
X}
 \frac{\partial f_{i}}{\partial Y}
-\frac{\partial\Phi}{\partial Y}\frac{\partial f_{i}}{\partial
X}\right) +\frac{e}{m_{i}}\frac{\omega_{ci}}{v_{\bot}}
 \left(\frac{\partial\Phi}{\partial \phi}
\frac{\partial f_{i}}{\partial v_{\bot}}
-\frac{\partial\Phi}{\partial v_{\bot}}
 \frac{\partial f_{i}}{\partial \phi} \right)\nonumber
\\ & \displaystyle -\frac{e}{m_{i}}\frac{\partial\Phi}{\partial z_{1}}
 \frac{\partial f_{i}}{\partial v_{z}}
=\frac{e}{m_{i}\omega_{ci}}\frac{\partial\Phi}{\partial Y}
\frac{\partial F
 _{i 0}}{\partial X}-\frac{e}{m_{i}}\frac{\omega_{ci}}{v_{\bot}}
\frac{\partial\Phi}{\partial \phi_{1}} \frac{\partial F
_{i0}}{\partial
 v_{\bot}} +\frac{e}{m_{i}}\frac{\partial\Phi}{\partial z_{1}}
  \frac{\partial F _{i 0}}{\partial v_{z}}.
\label{14}
\end{eqnarray}
This form of the Vlasov equation we consider as the most efficient
for the deriving the renormalized solution for $f_{i}$. Using the
system of equations for characteristics for Eq. (\ref{14}),
\begin{eqnarray}
& \displaystyle
dt=\frac{dX}{-\dfrac{e}{m_{i}\omega_{ci}}\dfrac{\partial\Phi}{\partial
Y_{1}}}=
\frac{dY}{\dfrac{e}{m_{i}\omega_{ci}}\dfrac{\partial\Phi}{\partial
X_{1}}}=
\frac{dv_{\bot}}{\dfrac{e}{m_{i}}\dfrac{\omega_{ci}}{v_{\bot}}
\dfrac{\partial\Phi}{\partial
\phi_{1}}}=\frac{d\phi_{1}}{-\dfrac{e}{m_{i}}\dfrac{\omega_{ci}}{v_{\bot}}
\dfrac{\partial\Phi}{\partial
v_{\bot}}}=\frac{dv_{z}}{-\dfrac{e}{m_{i}}\dfrac{\partial\Phi}{\partial
z_{1}}} \nonumber\\ & \displaystyle
=\frac{df_{i}}{\dfrac{e}{m_{i}\omega_{ci}}\dfrac{\partial\Phi}{\partial
Y} \dfrac{\partial F
 _{i0}}{\partial X}-\dfrac{e}{m_{i}}\dfrac{\omega_{ci}}{v_{\bot}}
\dfrac{\partial\Phi}{\partial \phi_{1}} \dfrac{\partial F
_{i0}}{\partial
 v_{\bot}} +\dfrac{e}{m_{i}}\dfrac{\partial\Phi}{\partial z_{1}}
  \dfrac{\partial F _{i0}}{\partial v_{z}}},\label{15}
\end{eqnarray}
the following nonlinear solution for the function $f_{i}$ with
known $F_{i0}$ is obtained :
\begin{eqnarray}
& \displaystyle f_{i} =
\frac{e}{m}\int\limits^{t}\left[\frac{1}{\omega_{ci}}\frac{\partial\Phi}{
\partial Y}\frac{\partial F_{i0}}{\partial X}-\frac{\omega_{ci}}{v_{\bot}}
\frac{\partial\Phi}{\partial \phi_{1}} \frac{\partial
F_{i0}}{\partial v_{\bot}} +\frac{\partial\Phi}{\partial z_{1}}
\frac{\partial F _{i0}}{\partial v_{z}} \right] dt'. \label{16}
\end{eqnarray}

Supposing  that the particle orbit disturbance $\delta X$ due to the
electrostatic plasma turbulence is sufficiently small, we find the
solution for system (\ref{15}). From the first equation of that
system we obtain
\begin{eqnarray}
X=\bar{X}+\delta X,  \qquad \delta
X=-\dfrac{e}{m_{i}\omega_{ci}}\int\limits^{t}\dfrac{\partial\Phi}
{\partial \bar{Y}}dt_{1},\label{17}
\end{eqnarray}
where $\bar{X}$ and $\bar{Y}$ are the guiding center coordinates
averaged over the turbulent pulsations. All other equations of
system (\ref{14}) have the following approximate solutions:
\begin{eqnarray}
Y=\bar{Y}+\delta Y,  \qquad \delta
Y=\dfrac{e}{m_{i}\omega_{ci}}\int\limits^{t}\dfrac{\partial\Phi}
{\partial \bar{X}}dt_{1},\label{18}
\end{eqnarray}
\begin{eqnarray}
v_{\bot}=\bar{v}_{\bot}+\delta v_{\bot},  \qquad \delta v_{\bot}=
\dfrac{e}{m_{i}}\dfrac{\omega_{ci}}{v_{\bot}}
\int\limits^{t}\dfrac{\partial\Phi}{\partial
\bar{\phi}}dt_{1},\label{19}
\end{eqnarray}
\begin{eqnarray}
 \phi=\bar{\phi}+\delta\phi,    \qquad
\delta\phi=-\dfrac{e}{m_{i}}\dfrac{\omega_{ci}}{v_{\bot}}
\int\limits^{t}\dfrac{\partial\Phi}{\partial
\bar{v}_{\bot}}dt_{1},\label{20}
\end{eqnarray}
\begin{eqnarray}
v_{z}=\bar{v}_{z}+\delta v_{z},   \qquad \delta v_{z}
=-\dfrac{e}{m}\int\limits^{t}_{0}\dfrac{\partial\Phi}{\partial
\bar{z}}dt_{1}.\label{21}
\end{eqnarray}
In Eqs.(\ref{17})-(\ref{21}) the perturbed electrostatic potential
$\Phi$ is defined in variables $\bar{X}$, $\bar{Y}$,
$\bar{v}_{\bot}$, $\bar{\phi}$, $\bar{v}_{z}$ and $\delta X$,
$\delta Y$, $\delta v_{\bot}$, $\delta\phi$, $\delta v_{z}$ as
\begin{eqnarray}
& \displaystyle \Phi\left(\mathbf{r},t
\right)=\sum_{n=-\infty}^{\infty}\int d\mathbf{k}J_{n}\left(
\frac{k_{\bot}\bar{v}_{\bot}}{\omega_{ci}}\right)\exp\Big[ik_{x}\bar{X}
+ik_{y}\bar{Y}+ik_{z}\bar{z} \nonumber\\ & \displaystyle
+ik_{z}\left(\bar{v}_{z}+ V_{i0}\left(\bar{X}
\right)\right)t-in\left(\bar{\phi}-\omega_{ci}t-\theta
\right)\Big]\exp\left(i\mathbf{k}\delta\mathbf{r}\left(t\right)\right)
\int d\omega \Phi\left(\mathbf{k},\omega \right)\exp\left(-i\omega
t\right) , \label{22}
\end{eqnarray}
where the perturbations of the ions orbits due to wave-ion
interactions,
\begin{eqnarray}
& \displaystyle \mathbf{k}\delta\mathbf{r}\left(t\right)=k_{x}\delta
X\left(t\right)+k_{y}\delta Y\left(t\right)+k_{z}\delta
z\left(t\right) -\frac{k_{\bot}\delta
v_{\bot}\left(t\right)}{\omega_{ci}}\sin \left(\phi-\theta\right)
-\frac{k_{\bot}\bar{v}_{\bot}}{\omega_{ci}}\cos
\left(\phi-\theta\right)\delta\phi\left(t\right), \label{23}
\end{eqnarray}
with
\begin{eqnarray}
& \displaystyle \delta z\left(t\right)=\int\limits^{t}\delta
v_{z}\left(t_{1} \right) dt_{1}+ V'_{0}\left(\bar{X}
\right)\int\limits^{t}\delta X\left(t_{1} \right)dt_{1}, \label{24}
\end{eqnarray}
are included. In Eq.(\ref{24}) the first term corresponds to the
scattering of ions along the magnetic field by the
$E_{z}=-ik_{z}\Phi$ projection of the turbulent electric field. The
second term corresponds to the combined effect of turbulence and
shear flow. This effect consists in turbulent scattering of ions
across the magnetic field along the velocity shear from the flow
layer with velocity $V_{0}\left(\bar{X} \right)$ into the layer with
velocity $V_{0}\left(\bar{X} \right)+V'_{0}\left(\bar{X}
\right)\delta X$ and enhancing by this means the transport of ions
along the shear flow. The order of value of the ratio of the second
term in Eq.(\ref{24}) to the first one is
$k_{y}V'_{0}/k_{z}\omega_{ci}$. In laboratory and ionospheric
plasmas\cite{Amatucci-1999} this parameter may exceed unity
considerably. In Eq.(\ref{24}) the terms of the second order in
$\delta X$, $\delta Y$, $\delta v_{\perp}$, $\delta v_{z}$ and
$\delta\phi$ are omitted. The Fourier transformed Poisson's
equation,
\begin{eqnarray}
& \displaystyle \Phi\left(\mathbf{k},\omega
\right)=-\frac{4\pi}{k^{2}}\sum_{\alpha=i,e}e_{\alpha}\delta
n_{\alpha}\left(\mathbf{k},\omega \right)
=-\frac{4\pi}{k^{2}}\sum_{\alpha=i,e}e_{\alpha}\int
f_{\alpha}d\mathbf{v},\label{25}
\end{eqnarray}
and Eq.(\ref{16}) in which the potential $\Phi\left(\mathbf{r},t
\right)$ is related with $\Phi\left(\mathbf{k},\omega \right)$ by
Eq.(\ref{22}) with $\delta\mathbf{r}$ determined by
Eqs.(\ref{17})-(\ref{21}), compose the system of nonlinear integral
equations for $\Phi\left(\mathbf{k},\omega \right)$ and
$f_{\alpha}$. We obtain from this system the nonlinear dispersion
equation which accounted for the scattering of ions by turbulence in
shear flow. We assume the equilibrium distribution function $F_{i0}$
to be a drifting Maxwellian
\begin{eqnarray}
& \displaystyle F_{i0}=\frac{n_{i0}\left(\bar{X} \right)}{\left(2\pi
v^{2}_{Ti}\right)^{3/2}} \exp\left(-\frac{v^{2}_{\bot}}{v^{2}_{Ti}}
-\frac{\left( v_{z}-V_{i0}\left(\bar{X} \right)\right)
^{2}}{v^{2}_{Ti}}\right). \label{26}
\end{eqnarray}
Then the ion density perturbation will be equal to
\begin{eqnarray}
& \displaystyle \delta n_{i}\left(\mathbf{k},\omega \right)= \int
d\mathbf{\bar{v}}f_{i}\left(\omega,\mathbf{k},\bar{v}_{\bot},\bar{\phi},\bar{v}_{z}
\right)\nonumber\\ &\displaystyle \nonumber=2\pi
i\frac{e}{T_{i}}\sum_{n=-\infty}^{\infty}\int\limits^{\infty}_{-\infty}
dv_{z}\int\limits^{\infty}_{0} dv_{\perp}v_{\perp}\int
d\omega_{1}\Phi\left(\mathbf{k},
\omega_{1}\right)\int\limits^{\infty}_{-\infty}dte^{i\left(\omega-\omega_{1}\right)t}
\int\limits^{\infty}_{0}dt_{1}F_{i0}\left(\bar{X},\bar{\mathbf{v}}\right)\nonumber\\
&\displaystyle\times\exp\Big[i\omega_{1}\left(t-t_{1}\right)-ik_{z}\left(\bar{v}_{z}+
V_{i0}\left(\bar{X}
\right)\right)\left(t-t_{1}\right)-in\omega_{ci}\left(t-t_{1}\right)
\Big]\nonumber\\ & \displaystyle
 \label{27}
\times\exp\Big[-i\mathbf{k}\Delta\mathbf{r}\left(t,
t_{1}\right)\Big]
J^{2}_{n}\left(\frac{k_{\bot}\bar{v}_{\bot}}{\omega_{ci}}\right)\left[k_{y}v_{di}+\frac{V'_
{i0}}{\omega_{ci}}k_{y} \bar{v}_{z}-n\omega_{ci}-k_{z}\bar{v}_{z}
\right]
\end{eqnarray}
where $\Delta\mathbf{r}\left(t,
t_{1}\right)=\delta\mathbf{r}\left(t\right)-\delta\mathbf{r}\left(t_{1}\right)$.
As in conventional renormalized theory\cite{Dum, Benford} we account
for the average effect of the perturbations of ions orbits in
Eq.(\ref{15}). For this we can use the cummulant
expansion\cite{Weinstock},
\begin{eqnarray}
& \displaystyle \Big\langle
\exp\left(-i\mathbf{k}\Delta\mathbf{r}\left(t,t_{1}\right)
\right)\Big\rangle=
\exp\left(\sum_{n=1}^{\infty}\frac{1}{n!}\Big\langle\Big[-i\mathbf{k}\Delta\mathbf{r}
\left(t,t_{1}\right)\Big]^{n}\Big\rangle_{c}\right),\label{28}
\end{eqnarray}
where $\langle...\rangle_{c}$ is the cummulant. In the following we
use the simplified approximation that the particles scattering by
plasma turbulence is a Gaussian process, for which expansion
(\ref{28}) under Markovian approximation is reduced to a single
term\cite{Dum},
\begin{eqnarray}
& \displaystyle\Big\langle
\exp\left(-i\mathbf{k}\Delta\mathbf{r}\left(t,t_{1}\right)
\right)\Big\rangle
\simeq\exp\left(-\frac{1}{2}\Big\langle\Big[\mathbf{k}\delta\mathbf{r}
\left(t-t_{1}\right)\Big]^{2}\Big\rangle\right).\label{29}
\end{eqnarray}
With these approximations the ion density perturbation will be equal
to
\begin{eqnarray}
& \displaystyle \delta n_{i}\left(\mathbf{k},\omega \right)= \int
d\mathbf{\bar{v}}f_{i}\left(\omega,\mathbf{k},\bar{v}_{\bot},\bar{\phi},\bar{v}_{z}
\right)\nonumber\\ &\displaystyle =
i\frac{e}{T_{i}}n_{i0}\left(\bar{X}\right)
\sum_{n=-\infty}^{\infty}A_{in}\left(k^{2}_{\perp}\rho^{2}_{i}\right)\left(k_{y}v_{di}-n\omega_{ci}
\right)
\Phi\left(\mathbf{k},\omega\right)R_{1}\left(\mathbf{k},\omega\right)\nonumber
\\ &\displaystyle -\frac{e}{T}n_{i0}\left(\bar{X}\right)
\sum_{n=-\infty}^{\infty}A_{in}\left(k^{2}_{\perp}\rho^{2}_{i}\right)
k^{2}_{z}v^{2}_{Ti}\left(1-\frac{k_{y}V'_{0}}{k_{z}\omega_{ci}}
\right)
\Phi\left(\mathbf{k},\omega\right)R_{2}\left(\mathbf{k},\omega\right),
\label{30}
\end{eqnarray}
where
\begin{eqnarray}
& \displaystyle
R_{1}\left(\mathbf{k},\omega\right)=\int\limits_{0}^{\infty}d\tau\exp
\left[i\Big(\omega-n\omega_{ci}-k_{z} V_{i0}\left(\bar{X}\right)
\Big)\tau-\frac{1}{2}k^{2}_{z}v^{2}_{Ti}\tau^{2}-\frac{1}{2}
\Big\langle
\big(\mathbf{k}\delta\mathbf{r}\left(\tau\right)\big)^{2}
\Big\rangle \right],\label{31}
\end{eqnarray}
\begin{eqnarray}
&\displaystyle
R_{2}\left(\mathbf{k},\omega\right)=\int\limits_{0}^{\infty}d\tau\tau\exp
\left[i\Big(\omega-n\omega_{c}-k_{z} V_{i0}\left(\bar{X}\right)
\Big)\tau-\frac{1}{2}k^{2}_{z}v^{2}_{Ti}\tau^{2}-\frac{1}{2}
\Big\langle
\big(\mathbf{k}\delta\mathbf{r}\left(\tau\right)\big)^{2}
\Big\rangle \right],\label{32}
\end{eqnarray}
with $\tau=t-t_{1}$ instead of $t_{1}$, are resonance functions
which replace the familiar resonant denominators of the linear
theory. Using Eq.(\ref{23}) and omitting the terms which oscillate
with frequency $n\omega_{ci}$ $(n=\pm1,\pm2,...)$ we can write the
mean square displacement as
\begin{eqnarray}
&\displaystyle
\frac{1}{2}\left\langle\left(\mathbf{k}\cdot\delta\mathbf{r}\right)^{2}\right\rangle=
\frac{1}{2}\left\langle\left(\mathbf{k}\cdot\delta\mathbf{r}\right)^{2}\right\rangle_{0}
+k_{x}k_{z}\left\langle\delta X\delta z\right\rangle
+k_{y}k_{z}\left\langle\delta Y\delta z\right\rangle
+\frac{1}{2}k^{2}_{z}\left\langle\left(\delta
z\right)^{2}\right\rangle , \label{33}
\end{eqnarray}
where
\begin{eqnarray}
&\displaystyle
\left\langle\left(\mathbf{k}\cdot\delta\mathbf{r}\right)^{2}\right\rangle_{0}=
k^{2}_{x} \left\langle\left(\delta
X\right)^{2}\right\rangle+k^{2}_{y}\left\langle\left(\delta
Y\right)^{2}\right\rangle+2k_{x}k_{y} \left\langle\delta X\delta
Y\right\rangle +\frac{1}{2}\frac{k^{2}_{\perp}v^{2}_{\perp}}
{\omega^{2}_{c}}\left\langle \left(\delta
\phi\right)^{2}\right\rangle+\frac{1}{2}\frac{k^{2}_{\perp}}
{\omega^{2}_{c}}\left\langle \left(\delta
v_{\perp}\right)^{2}\right\rangle.\label{34}
\end{eqnarray}
The term
$\left\langle\left(\mathbf{k}\cdot\delta\mathbf{r}\right)^{2}\right\rangle_{0}$
exists in plasma even without shear flow\cite{Dum}. The first three
terms in Eq.(\ref{33}) describe the diffusion of the ion guiding
center coordinates $X$ and $Y$ and the last two terms describe the
random changes of the IC radius $v_{\perp}/\omega_{ci}$ and phase
angle $\phi$ of the Larmor orbit. In the asymptotic limit of large
time compared with the correlation time $\tau_{corr}$ of the
turbulent electric field along the particle orbit, and after
averaging over ions velocity assuming a Maxwellian distribution we
obtain that the term
$\left\langle\left(\mathbf{k}\cdot\delta\mathbf{r}\right)^{2}\right\rangle_{0}$
is equal to
\begin{eqnarray}
&\displaystyle \frac{1}{2}\left\langle\left(\mathbf{k}\cdot\delta
\mathbf{r}\right)^{2}\right\rangle_{0}=C_{1}t=
\frac{e^{2}t}{2m_{i}^{2}\omega_{ci}^{2}}\mathrm{Re}
\sum_{n_{1}=-\infty}^{\infty}\int
d\mathbf{k}_{1}|\Phi\left(\mathbf{k}_{1}\right)|^{2}
e^{-k^{2}_{\bot}\rho^{2}_{i}}\left[ 2\left(k_{x}k_{1y}-k_{1x}k_{y}
\right)^{2}I_{n_{1}}\left(k^{2}_{\bot}\rho^{2}_{i}\right)\right.
\nonumber  \\ &\displaystyle
\left.+\frac{k^{2}_{\bot}k^{2}_{1\bot}}{2}
\Big(I_{n_{1}+1}\left(k^{2}_{\bot}\rho^{2}_{i}
\right)+I_{n_{1}-1}\left(k^{2}_{\bot}\rho^{2}_{i}
\right)\Big)\right]\nonumber  \\ &\displaystyle
\times\int\limits_{0}^{ \infty } d\tau
\exp\left[-i\left(\omega-n\omega_{c}-k_{z}
V_{i0}\left(\bar{X}\right)\right)\tau-\frac{1}{2} \Big\langle
\big(\mathbf{k}_{1}\delta\mathbf{r}\left(\tau\right)\big)^{2}
\Big\rangle\right], \label{35}
\end{eqnarray}
Last three terms in Eq.(\ref{33}), which are due to scattering of
ions along the magnetic field, are equal to
\begin{eqnarray}
& \displaystyle k_{x}k_{z}\left\langle\delta X\delta z\right\rangle
+k_{y}k_{z}\left\langle\delta Y\delta z\right\rangle
+\frac{1}{2}k^{2}_{z}\left\langle\left(\delta
z\right)^{2}\right\rangle =C_{2}t^{2}+ C_{3}t^{3} , \label{36}
\end{eqnarray}
with
\begin{eqnarray}
& \displaystyle C_{2}=
\frac{e^{2}}{2m_{i}^{2}\omega_{ci}}\mathrm{Re}
\sum_{n_{1}=-\infty}^{\infty}\int
d\mathbf{k}_{1}|\Phi\left(\mathbf{k}_{1}\right)|^{2}
e^{-k^{2}_{\bot}\rho^{2}_{i}}
k_{z}k_{1z}\left(1+\frac{k_{1y}V'_{i0}}{k_{1z}
\omega_{ci}}\right)\left(k_{x}k_{1y}-k_{1x}k_{y} \right) \nonumber\\
&\displaystyle \times I_{n_{1}}\left(k^{2}_{\bot}\rho^{2}_{i}
\right)\int\limits_{0}^{ \infty } d\tau
\exp\left[-i\left(\omega-n\omega_{c}-k_{z}
V_{i0}\left(\bar{X}\right)\right)\tau-\frac{1}{2} \Big\langle
\big(\mathbf{k}_{1}\delta\mathbf{r}\left(\tau\right)\big)^{2}
\Big\rangle\right], \label{37}
\end{eqnarray}
\begin{eqnarray}
& \displaystyle C_{3}=\frac{e^{2}}{3m_{i}^{2}}\mathrm{Re}
\sum_{n_{1}=-\infty}^{\infty}\int
d\mathbf{k}_{1}|\Phi\left(\mathbf{k}_{1}\right)|^{2}
e^{-k^{2}_{\bot}\rho^{2}_{i}}k_{z}^{2}k_{1z}^{2}\left(1+\frac{k_{1y}V'_{i0}}
{k_{1z}\omega_{ci}}\right)^{2}\nonumber\\ &\displaystyle \times
I_{n_{1}}\left(k^{2}_{\bot}\rho^{2}_{i} \right)\int\limits_{0}^{
\infty } d\tau \exp\left[-i\left(\omega-n\omega_{c}-k_{z}
V_{i0}\left(\bar{X}\right)\right)\tau-\frac{1}{2} \Big\langle
\big(\mathbf{k}_{1}\delta\mathbf{r}\left(\tau\right)\big)^{2}
\Big\rangle\right]. \label{38}
\end{eqnarray}
where the term with $C_{2}$ originates from the first two terms on
the left of Eq.(\ref{36}), whereas the term with $C_{3}$ resulted
from the last left term of Eq.(\ref{36}). Due to the term
$V'_{0}\left(\bar{X} \right)\int\limits^{t}\delta X\left(t_{1}
\right)dt_{1}$ in Eq.(\ref{24}), term $C_{3}$ includes the effect of
the anomalous viscosity, resulted from the turbulent redistribution
of the shear flow moment.

The frequencies and growth rates of ion cyclotron instabilities,
which were considered in Ref.\cite{Mikhailenko2006}, are determined
under condition of weak IC damping, i.e. for
$|\omega-k_{z}V_{i0}\left(\bar{X}\right)-n\omega_{ci}|/k_{z}v_{Ti}\gg
1$. Under this condition the asymptotics of integrals (\ref{31}) and
(\ref{32}) are determined by the contributions over the time
interval $0<\tau\ll
|\omega-k_{z}V_{i0}\left(\bar{X}\right)-n\omega_{ci}|^{-1}$ and
contributions from the stationary phase point. Our calculations of
the integrals $R_{1}$ and $R_{2}$ give that the contribution from
the stationary phase point is negligible for both integrals and main
contributions proceed from the vicinity of $\tau=0$ point and are
equal to
\begin{eqnarray}
& R_{1}=\displaystyle \int\limits_{0}^{\infty} d\tau
\exp\Big(i\delta\omega\tau-\mathbb{C}_{2}\tau^{2}-C_{3}\tau^{3}
\Big)\approx\frac{i}{\delta\omega}
+\frac{2i\mathbb{C}_{2}}{\left(\delta\omega\right)^{3}}-
\frac{6C_{3}}{\left(\delta\omega\right)^{4}}, \label{39}
\end{eqnarray}
\begin{eqnarray}
& R_{2}=\displaystyle \int\limits_{0}^{\infty} d\tau\tau
\exp\Big(i\delta\omega\tau-\mathbb{C}_{2}\tau^{2}-C_{3}\tau^{3}
\Big)\approx-\frac{1}{\left(\delta\omega\right)^{2}}
-\frac{6\mathbb{C}_{2}}{\left(\delta\omega\right)^{4}}-
\frac{24iC_{3}}{\left(\delta\omega\right)^{5}}, \label{40}
\end{eqnarray}
where now  $\delta\omega\left(\mathbf{k}\right)$ is the renormalized
version of the linear $\delta\omega\left(\mathbf{k}\right)$,
presented in Introduction,
\begin{eqnarray}
& \displaystyle
\delta\omega\left(\mathbf{k}\right)=\omega-k_{z}V_{i0}\left(\bar{X}\right)-n\omega_{ci}+iC_{1},
\label{41}
\end{eqnarray}
and $\mathbb{C}_{2}=C_{2}-\frac{1}{2}k_{z}^{2}v^{2}_{Ti}$. In
Eq.(\ref{39}) the first term $i/\delta\omega$ includes known
\cite{Dum} ion cyclotron resonance broadening effect; the following
two ones, which are proportional to $C_{2}$ and $C_{3}$, reflect the
combined effect of turbulent scattering and shear flow. Using above
obtained results for $R_{1}$ and $R_{2}$ in Eq.(\ref{30}) we obtain
the renormalized version of the equation for perturbed ion density,
$\delta n_{i}\left(\mathbf{k},\omega\right)$, which accounted for
the the combined effect of shear flow and ions scattering by
electrostatic turbulence,
\begin{eqnarray}
& \displaystyle \delta n_{i}\left(\mathbf{k},\omega\right)=
-\frac{e_{i}n_{0i}}{T_{i}}\Phi\left(\mathbf{k},\omega\right)\left[1-
\sum_{n=-\infty}^{\infty}\left(\frac{\omega-k_{z}V_{0}\left(\bar{X}\right)-k_{y}
v_{di}+iC_{1}}{\delta\omega}\right) \right. \nonumber\\
&\displaystyle \left.
\times\left(1+\frac{k_{z}^{2}v_{Ti}^{2}}{\left(\delta\omega\right)^{2}}\right)
A_{in}\left(k^{2}_{\perp}\rho^{2}_{i}\right)
\right]-\frac{e_{i}n_{0i}}{T_{i}}\Phi\left(\mathbf{k},\omega\right)
\frac{k_{y}V'_{0}}{k_{z}\omega_{ci}}\sum_{n=-\infty}^{\infty}
\frac{k_{z}^{2}v_{Ti}^{2}}{\left(\delta\omega\right)^{2}}
A_{in}\left(k^{2}_{\perp}\rho^{2}_{i}\right) \nonumber\\
&\displaystyle
-\frac{2e_{i}n_{0i}}{T_{i}}\Phi\left(\mathbf{k},\omega\right)
\sum_{n=-\infty}^{\infty}\frac{\left(-n\omega_{ci}+k_{y}v_{di}
\right)
}{\left(\delta\omega\right)^{3}}A_{in}\left(k^{2}_{\perp}\rho^{2}_{i}\right)\left(C_{2}+3i\frac{C_{3}}{
\delta\omega}\right) \label{42}\\ &\displaystyle +
 6\frac{e_{i}n_{0i}}{T_{i}}\Phi\left(\mathbf{k},\omega\right)
\left(1- \frac{k_{y}V'_{0}}{k_{z}\omega_{ci}}\right)
\sum_{n=-\infty}^{\infty}
\frac{k_{z}^{2}v_{Ti}^{2}}{\left(\delta\omega\right)^{4}}
A_{in}\left(k^{2}_{\perp}\rho^{2}_{i}\right)
\left(C_{2}+4i\frac{C_{3}}{\delta\omega}\right).\nonumber
\end{eqnarray}
Inserting Eq.(\ref{42}) to Poisson's equation (\ref{25}), we obtain
the  nonlinear dispersion equation
\begin{eqnarray}
& \displaystyle
1+\varepsilon_{0i}\left(\mathbf{k},\omega\right)+\varepsilon_{e}\left(\mathbf{k},\omega\right)
+\varepsilon_{sh}\left(\mathbf{k},\omega\right)=0,\label{43}
\end{eqnarray}
which together with Eqs.(\ref{35})-(\ref{38}) comprises the system
of nonlinear integral equations for spectral intensity
$|\Phi\left(\mathbf{k}_{1}\right)|^{2}$ and frequency $\omega$. This
system is derived under condition that the dominant nonlinear effect
is a randomization of ion orbits. This effect enters the dispersion
equation only through resonance functions (\ref{39}) and(\ref{40}).
In Eq.(\ref{43}) we introduce the notations
\begin{eqnarray}
& \displaystyle
\varepsilon_{0i}\left(\mathbf{k},\omega\right)=\text{Re}\,
\varepsilon^{\left(1\right)}_{i}\left(\mathbf{k},\omega\right)
+\text{Re}\varepsilon^{\left(2\right)}_{i}\left(\mathbf{k},\omega\right)\nonumber\\
&\displaystyle=\frac{1}{k^{2}\lambda^{2}_{Di}}\left\{ 1-
\sum_{n=-\infty}^{\infty}\left(\frac{\omega-k_{z}V_{0}\left(\bar{X}\right)-k_{y}
v_{di}+iC_{1}}{\delta\omega}\right)\left(1+\frac{k_{z}^{2}v_{Ti}^{2}}{\left(\delta\omega\right)^{2}}\right)
A_{in}\left(k^{2}_{\perp}\rho^{2}_{i}\right) \right\}\nonumber\\
&\displaystyle
+\frac{1}{k^{2}\lambda^{2}_{Di}}\frac{k_{y}V'_{0}}{k_{z}\omega_{ci}}\sum_{
n=-\infty}^{\infty}\frac{k_{z}^{2}v_{Ti}^{2}}{\left(\delta\omega\right)^{2}}
A_{in}\left(k^{2}_{\perp}\rho^{2}_{i}\right)  ,\label{44}
\end{eqnarray}
\begin{eqnarray}
& \displaystyle
\varepsilon_{e}\left(\mathbf{k},\omega\right)=\frac{1}{k^{2}\lambda
_{De}^{2} }\left[1 + i\sqrt {\frac{\pi
}{2}}z_{0e}\exp\left(-z^{2}_{e0}\right)\right]=\frac{1}{k^{2}\lambda
_{De}^{2}
}+\triangle\varepsilon_{e}\left(\mathbf{k},\omega\right),\label{45}
\end{eqnarray}
where $z_{e0}=\left(\omega-k_{z}V_{e0}\right)/\sqrt{2}k_{z}v_{Te}$
and
\begin{eqnarray}
& \displaystyle  \varepsilon_{sh}\left(\mathbf{k},\omega\right)=
\frac{2}{k^{2}\lambda^{2}_{Di}}\sum_{n=-\infty}^{\infty}\frac{\left(-n\omega_{
ci}+k_{y}v_{di} \right)
}{\left(\delta\omega\right)^{3}}A_{in}\left(k^{2}_{\perp}\rho^{2}_{i}\right)\left(C_{2}+3i\frac{C_{3}}{
\delta\omega}\right)\nonumber\\ &\displaystyle
-\frac{6}{k^{2}\lambda^{2}_{Di}}\left(1-
\frac{k_{y}V'_{0}}{k_{z}\omega_{ci}}\right)
\sum_{n=-\infty}^{\infty}
\frac{k_{z}^{2}v_{Ti}^{2}}{\left(\delta\omega\right)^{4}}
A_{in}\left(k^{2}_{\perp}\rho^{2}_{i}\right)
\left(C_{2}+4i\frac{C_{3}}{\delta\omega}\right). \label{46}
\end{eqnarray}
The system of equations (\ref{35}), (\ref{37}), (\ref{38}) and
(\ref{43}) is the main result of this theory. To obtain the solution
$\omega\left(\mathbf{k}\right)$ of Eq.(\ref{43}) we expand it about
the solution $\omega_{0}\left(\mathbf{k}\right)$ of the linear
dispersion equation,
$1+\varepsilon_{0i}\left(\bf{k},\omega_{0}\left(\mathbf{k}\right)
\right)+1/k^{2}\lambda_{De}^{2}=0$,
\begin{eqnarray}
&\displaystyle
1+\varepsilon_{0i}\left(\mathbf{k},\omega_{0}\left(\mathbf{k}\right)\right)+
\frac{\partial\varepsilon_{0i}\left(\mathbf{k},\omega_{0}\right)}{\partial\omega_{0}}\left(\omega+iC_{1}
-\omega_{0}\left(\mathbf{k}\right)\right)+\triangle\varepsilon_{e}\left(\mathbf{k},\omega\right)
+\varepsilon_{sh}\left(\mathbf{k},\omega_{0}\left(\mathbf{k}\right)\right)=0.\label{47}
\end{eqnarray}
The solution is
\begin{eqnarray}
&\displaystyle
\omega\left(\mathbf{k}\right)=\omega_{0}\left(\mathbf{k}\right)-iC_{1}-
\frac{\triangle\varepsilon_{e}\left(\mathbf{k},\omega\right)+\varepsilon_{sh}
\left(\mathbf{k},\omega_{0}\left(\mathbf{k}\right)\right)}
{\dfrac{\partial\varepsilon_{0i}\left(\mathbf{k},\omega_{0}\right)}{\partial\omega_{0}}}.\label{48}
\end{eqnarray}
Thus, the nonlinear growth/damping rate, which accounts for the
scattering of ions by plasma turbulence in along-field shear flow is
equal to
\begin{eqnarray}
&\displaystyle
\gamma\left(\mathbf{k}\right)=\text{Im}\,\omega\left(\mathbf{k}\right)
=\gamma_{0}\left(\mathbf{k}\right)-C_{1}+\gamma_{sh}\left(\mathbf{k}\right),\label{49}
\end{eqnarray}
where
\begin{eqnarray}
&\displaystyle
\gamma_{0}\left(\mathbf{k}\right)=-\frac{\text{Im}\,\triangle\varepsilon_{e}\left(\mathbf{k},\omega_{0}
\left(\mathbf{k}\right)\right)}
{\dfrac{\partial\varepsilon_{0i}\left(\mathbf{k},\omega_{0}\right)}{\partial\omega_{0}}}\,,\qquad
\gamma_{sh}\left(\mathbf{k}\right)=-\frac{\text{Im}\,\varepsilon_{sh}\left(\mathbf{k},\omega_{0}
\left(\mathbf{k}\right)\right)}
{\dfrac{\partial\varepsilon_{0i}\left(\mathbf{k},\omega_{0}\right)}{\partial\omega_{0}}}\,.\label{50}
\end{eqnarray}
In the case of the shearless plasma flow IC resonance broadening
effect, determined by $C_{1}$, leads to the saturation of the
current driven IC instability\cite{Dum}. Below we examine the role
of the shear flow driven term $\gamma_{sh}\left(\mathbf{k}\right)$
in the nonlinear evolution of the shear flow modified and shear flow
driven kinetic IC instabilities. It follows from Eq.(\ref{50}) that
when $\gamma_{sh}\left(\mathbf{k}\right)<0 $ the combined effect of
the shear flow and IC resonance broadening leads to the steady
state, $\gamma\left(\mathbf{k}\right)=0$, and the saturation level
of the IC turbulence is estimated from the equation
\begin{eqnarray}
&\displaystyle \gamma_{0}\left(\mathbf{k}\right)=C_{1}
-\gamma_{sh}\left(\mathbf{k}\right).\label{51}
\end{eqnarray}
However, when $\gamma_{sh}\left(\mathbf{k}\right)>0$, a nonlinear
instability may occur.

\section*{III. NONLINEAR STAGE AND SATURATION OF THE ION CYCLOTRON INSTABILITIES
OF MAGNETIC FIELD-ALIGNED PLASMA SHEAR FLOW}

In this section we apply the renormalized dispersion equation
(\ref{43}) to the qualitative analysis of the nonlinear evolution of
the  kinetic IC instabilities presented in Introduction, which are
developed due to inverse electron Landau damping. It follows from
Eqs.(\ref{6}) and (\ref{10}), that growth rates
$\gamma_{0(1,2)}\simeq\gamma_{e(1,2)}$ of the investigated
instabilities, attain theirs maximal values at
$|z_{e}|=|n\omega_{ci}-k_{z}\left(V_{e0}-V_{i0}\right)|/\sqrt{2}|k_{z}|v_{Te}\lesssim
1$. For $n\omega_{ci}\sim |k_{z}\left(V_{e0}-V_{i0}\right)|$ that
gives the estimate
\begin{eqnarray}
&\displaystyle
\frac{k_{z}}{k_{\perp}}\sim\frac{v_{Ti}}{v_{Te}}\frac{1}{k_{\perp}\rho_{i}},
\label{52}
\end{eqnarray}
which links possible values of $k_{z}/k_{\perp}$ and
$k_{\perp}\rho_{i}$ for which the maximum values of the growth rates
may be developed. Applying this estimate to condition (\ref{3}) of
weak flow shear, and to the condition of strong flow shear, opposite
to (\ref{3}), we come to the following estimates:
\begin{eqnarray}
&\displaystyle \left|\frac{V'_{0}}{\omega_{ci}}\right|
<\frac{v_{Te}}{v_{Ti}}\frac{A_{in}\left(k^{2}_{\perp}\rho^{2}_{i}\right)}
{k_{\perp}\rho_{i}\left(1-A_{i0}\left(k^{2}_{\perp}\rho^{2}_{i}\right)+\tau
\right)}\label{53}
\end{eqnarray}
for weak flow shear, and
\begin{eqnarray}
&\displaystyle \left|\frac{V'_{0}}{\omega_{ci}}\right|
>\frac{v_{Te}}{v_{Ti}}\frac{A_{in}\left(k^{2}_{\perp}\rho^{2}_{i}\right)}{k_{\perp}\rho_{i}
\left(1-A_{i0}\left(k^{2}_{\perp}\rho^{2}_{i}\right)+\tau
\right)}\label{54}
\end{eqnarray}
for strong flow shear, which determine the possible values for
$k_{\perp}\rho_{i}$ in the cases of weak or strong flow shear for
which maximal values of the growth rates may be attained. Accounting
for these estimates, as well as the results of the linear theory
presented in Introduction, we consider the cases of plasma flows
with weak and strong flow shear separately.

\subsection*{A.Weak flow shear}
In this subsection we consider the case of weak flow shear, limited
by condition (\ref{3}), for which two kinetic IC instabilities may
develop \cite{Mikhailenko2006}. The first one is the shear flow
modified current driven IC instability, which is excited when
$k_{z}\left(V_{0e}-V_{0i}\right)+k_{y}v_{de}>n\omega_{ci}$ with
frequency $\omega_{01}=n\omega_{ci}-k_{z}V_{0i}+\delta\omega_{01}$
and with growth rate $\gamma_{01}=\gamma_{1i}+\gamma_{1e}$, where
$\delta\omega_{01}$,  $\gamma_{1i}$ and $\gamma_{1e}$ are determined
by Eqs.(\ref{4}), (\ref{5}), (\ref{6}), respectively. The second one
is the shear flow driven IC instability\cite{Mikhailenko2006}. It is
excited with frequency
$\omega_{02}=n\omega_{ci}-k_{z}V_{0i}+\delta\omega_{02}$ and with
growth rate $\gamma_{02}=\gamma_{2i}+\gamma_{2e}$, where
$\delta\omega_{02}$,  $\gamma_{2i}$ and $\gamma_{2e}$ are determined
by Eqs.(\ref{8}), (\ref{9}), (\ref{10}), respectively, when the
velocity of the relative drift between ions and electrons is below
the critical value $V_{0e}^{\left(c\right)}$, roughly estimated as
$V_{0e}^{\left(c\right)}= V_{0i}$+
$\left(n\omega_{ci}-k_{y}v_{de}\right)/k_{z}$, i.e. under conditions
at which  modified by  shear flow current driven ion cyclotron
instability does not develop. In the case of shearless current the
dominant nonlinear effect responsible for the saturation of the
current driven IC instability is the turbulent broadening of the ion
cyclotron resonances\cite{Dum, Benford}, determined by the term
$C_{1}$ in Eq.(\ref{49}). The shear flow introduces new combined
effect of shear flow and turbulent scattering, which is incorporated
in $\gamma_{sh}$ term in Eq.(\ref{49}). Here we evaluate the joint
effect of shear flow and IC turbulence, determined by the terms
$C_{1}$ and $\gamma_{sh}$ in (\ref{49}) on the nonlinear evolution
of the above mentioned shear flow modified and shear flow driven
kinetic IC instabilities.

\subsection*{1. Shear flow modified ion cyclotron current driven instability}
Under the condition of the weak flow shear (\ref{3}), the ratio
$\left|\gamma_{01}C_{2}/C_{3}\right|$ of the $C_{2}$ and $C_{3}$
contained terms in $\gamma_{sh}$ is
\begin{eqnarray}
&\displaystyle
\left|\frac{\gamma_{01}C_{2}}{C_{3}}\right|\sim\frac{\gamma_{01}}{\omega_{ci}}\frac{k_{y}^{2}}{k_{z}^{2}}
\frac{1}{\left(1+\dfrac{k_{y}V'_{i0}}{k_{z}\omega_{ci}}\right)}>\frac{k_{y}^{2}\rho_{i}^{2}}
{A_{in}\left(k^{2}_{\perp}\rho^{2}_{i}\right)}\left(1-A_{i0}\left(k^{2}_{\perp}\rho^{2}_{i}\right)
+\tau\right)^{2}.\label{55}
\end{eqnarray}
It follows from Eq.(\ref{55}) that for $k_{\bot}\rho_{i}\gg 1$
$\left|\gamma_{01}C_{2}/C_{3}\right|>k^{4}_{y}\rho^{4}_{i}\gg 1$ and
for $k_{\bot}\rho_{i}\ll 1$
$\left|\gamma_{01}C_{2}/C_{3}\right|>\left(k_{y}\rho_{i}\right)^{2-4n}\gg
1$. For this reason, for $k_{\bot}\rho_{i}\gg 1$ as well as for
$k_{\bot}\rho_{i}\ll 1$, term with $C_{3}$ in Eq.(\ref{45}) may be
omitted and $\gamma_{sh}$ is determined by $C_{2}$ term in the first
line of Eq.(\ref{46}) as
\begin{eqnarray}
&\displaystyle
\gamma_{sh\left(1\right)}\left(\mathbf{k}\right)\simeq-\frac{6}{\left(1-A_{i0}\left(k^{2}_{\perp}\rho^{2}_{i}\right)
+\tau\right)}\frac{n\omega_{ci}\gamma_{0}C_{2}}{\left(\delta\omega_{01}\right)^{3}}
A_{in}\left(k^{2}_{\perp}\rho^{2}_{i}\right).\label{56}
\end{eqnarray}
Note, that because of $k_{z}v_{Ti}\ll \delta\omega_{01}$, the second
line in Eq.(\ref{46}) may be neglected. Now we estimate the ratio
$C_{1}/\gamma_{sh\left(1\right)}$ under condition of weak flow
shear. We have that
\begin{eqnarray}
&\displaystyle
\frac{C_{1}}{\gamma_{sh\left(1\right)}}>\frac{k_{y}^{2}\rho_{i}^{2}}
{A_{in}\left(k^{2}_{\perp}\rho^{2}_{i}\right)}\left(1-A_{i0}\left(k^{2}_{\perp}\rho^{2}_{i}\right)
+\tau\right)^{2}\gg 1 \label{57}
\end{eqnarray}
for $k_{\bot}\rho_{i}\gg 1$ as well as for  $k_{\bot}\rho_{i}\ll 1$.
Therefore the effect of shear flow on the saturation of the shear
flow modified IC current driven instability is subdominant. The
level of the IC turbulence in the steady state,
$e\widetilde{\Phi}/T_{e}$, where
$\widetilde{\Phi}=\left(\int|\Phi\left(\mathbf{k}_{1}\right)|^{2}d\mathbf{k}_{1}\right)^{1/2}$
is the root-mean-square (rms) magnitude of the perturbed
electrostatic potential, is estimated from the balance equation
$\gamma_{0}\left(\mathbf{k}\right)=C_{1}$. Using the  mean value
theorem for the integral over $\mathbf{k}_{1}$ in Eq.(\ref{35}) we
have
\begin{eqnarray}
\frac{e\widetilde{\Phi}}{T_{i}}\sim\frac{1}{\left(k_{\bot}\rho_{i}\right)^{5/2}}\gtrsim
\left(\frac{v_{Ti}}{v_{Te}}\frac{V'_{i0}}{\omega_{ci}}\right)^{5/4}
\label{58}
\end{eqnarray}
in the short wavelength, $k_{\bot}\rho_{i}\gg 1$, part of the
spectrum, and
\begin{eqnarray}
\frac{e\widetilde{\Phi}}{T_{i}}\sim
2^{-\frac{n+1}{2}}\left(k_{\bot}\rho_{i}\right)^{n-1}\gtrsim
2^{-\frac{n+1}{2}}\left(\frac{v_{Ti}}{v_{Te}}\frac{V'_{i0}}{\omega_{ci}}n!\right)^{\frac{n-1}{2n-1}}
\label{59}
\end{eqnarray}
in the long wavelength part, $k_{\bot}\rho_{i}\ll 1$, where we have
used Eqs.(\ref{52}) and (\ref{53}) in limits $k_{\bot}\rho_{i}\gg 1$
and $k_{\bot}\rho_{i}\ll 1$, respectively. It follows from estimates
(\ref{58}) and (\ref{59}) that $\widetilde{\Phi}$ has a very low
level (\ref{58}) for IC turbulence in the short wavelength part of
spectrum and attains its maximal value for $n=1$ and
$k_{\bot}\rho_{i}\lesssim 1$ \cite{Dum}, which is equal to
\begin{eqnarray} \frac{e\widetilde{\Phi}}{T_{i}}\sim \frac{1}{2}\,. \label{60}
\end{eqnarray}

\subsection*{2. Shear flow driven ion cyclotron instability}
Now we estimate the ratio $\gamma_{02}C_{2}/C_{3}$ which determines
relative importance of the $C_{2}$ and $C_{3}$ contained terms in
$\text{Im}\,\varepsilon_{sh}\left(\mathbf{k},\omega_{02}\left(\mathbf{k}\right)\right)$.
It follows from Eqs.(\ref{7}), (\ref{9}), (\ref{37}) and (\ref{38})
that
\begin{eqnarray}
&\displaystyle
\left|\frac{\gamma_{02}C_{2}}{C_{3}}\right|\sim\tau\left|\frac{k_{y}}{k_{z}}\frac{V'_{i0}}{\omega_{ci}}\right|
\frac{k^{4}_{z}\rho^{4}_{i}}{A_{in}\left(k^{2}_{\perp}\rho^{2}_{i}\right)}
\frac{k^{ 2}_{y}}{k^{2}_{z}}<
\frac{k^{2}_{y}\rho^{2}_{i}}{\left(1-A_{i0}\left(k^{2}_{\perp}\rho^{2}_{i}\right)
+\tau+k^{2}\lambda^{2}_{Di}\right)}, \label{61}
\end{eqnarray}
where constraint (\ref{3}) was used. Therefore for long wavelength,
$k_{\perp}\rho_{i}<1$, IC mode $\omega_{2}\left(\mathbf{k}\right)$
the term with $C_{3}$ is dominant in $\text{Im}\,\varepsilon_{sh}$.
In this case $\gamma^{\left(2\right)}_{sh}$ is equal to
\begin{eqnarray}
&\displaystyle \gamma^{\left(2\right)}_{sh}\simeq
A_{in}\left(k_{1\bot}^{2}\rho_{i}^{2}
\right)C_{3}\left(\frac{k_{y}V'_{i0}}
{k_{z}\omega_{ci}}\right)^{2}\frac{\left(n\omega_{ci}
+24\dfrac{k^{2}_{z}v^{2}_{Ti}}{\delta\omega_{02}}\left(1-\dfrac{k_{y}V'_{i0}}
{k_{z}\omega_{ci}}\right)\right)}{\left(1+\tau\right)\left(\delta\omega_{02}\right)^{2}\delta\omega_{01}}\label{62},
\end{eqnarray}
where the approximation of the separate IC mode was used. In
Eq.(\ref{62}) $C_{3}$ is determined by the relation
\begin{eqnarray}
&\displaystyle
C_{3}+\frac{2e^{2}}{m^{2}_{i}}\sum_{n_{1}=-\infty}^{\infty}\int
d\mathbf{k}_{1}|\Phi\left(\mathbf{k}_{1}\right)|^{2}
A_{in}\left(k_{1\bot} ^{2} \rho _{i}^{2}
\right)k_{z}^{2}k_{1z}^{2}\left(\frac{k_{1y}V'_{i0}}
{k_{1z}\omega_{ci}}\right)^{2}
\frac{C_{3}}{\left(\delta\omega_{02}\right)^{4}}\nonumber\\
&\displaystyle=\frac{e^{2}}{3m^{2}_{i}}\sum_{n_{1}=-\infty}^{\infty}\int
d\mathbf{k}_{1}|\Phi\left(\mathbf{k}_{1}\right)|^{2}
A_{in}\left(k_{1\bot} ^{2} \rho _{i}^{2 }
\right)k_{z}^{2}k_{1z}^{2}\left(\frac{k_{1y}V'_{i0}}
{k_{1z}\omega_{ci}}\right)^{2}\frac{\gamma_{02}}{\left(\delta\omega_{02}\right)^{2}},\label{63}
\end{eqnarray}
which follows from (\ref{38}) and (\ref{39}). In Eq.(\ref{63}) the
first term on the left side, $C_{3}$, is dominant over the second
one when
\begin{eqnarray}
&\displaystyle\frac{e\widetilde{\Phi}}{T_{i}}\lesssim\frac{k^{2}_{z}v_{Ti}^{2}}{\omega_{ci}^{2}}
\frac{k_{y}V'_{i0}}{k_{z}\omega_{ci}}\frac{1}{A^{1/2}_{in}\left(k_{\bot}^{2}\rho_{i}^{2}
\right)}.\label{64}
\end{eqnarray}
However at level (\ref{64}) $\gamma^{\left(2\right)}_{sh}\sim
24\left(k_{y}V'_{0i}/k_{z}\omega_{ci}\right)^{2}\gamma_{02}\gg
\gamma_{02}$, i.e. the saturation occurs before the second term in
left hand side of Eq.(\ref{63}) becomes comparable with the first
one. In the case $k_{\perp}\rho_{i}<1$
\begin{eqnarray}
&\displaystyle \frac{C_{1}}{ \gamma^{\left(2\right)}_{sh}}\sim
k_{\bot} ^{2} \rho _{i}^{2}
\frac{\left(1+\tau\right)}{\tau}\left(\frac{k_{z}\omega_{ci}}{k_{y}V'_{0i}}\right)^{2}\ll1.
\label{65}
\end{eqnarray}
Thus the nonlinear damping rate $\gamma^{\left(2\right)}_{sh}$
(\ref{62}), which for $k_{\perp}\rho_{i}<1$ may be approximated as
\begin{eqnarray}
&\displaystyle
\gamma^{\left(2\right)}_{sh}\simeq-\frac{24\left(\tau+k_{\bot}^{2}
\rho_{i}^{2}\right)C_{3}}{1+\tau}
\frac{\omega_{ci}^{2}}{k^{4}_{z}v^{4}_{Ti}}, \label{66}
\end{eqnarray}
dominates over $C_{1}$ in this case. The level of the IC turbulence
in this case will be determined by the balance equation
$\gamma^{\left(2\right)}_{sh}\simeq\gamma_{02}$ with $C_{3}$ term
determined by Eq.(\ref{63}) without the second term. This balance
equation gives the level
\begin{eqnarray}
&\displaystyle\frac{e\widetilde{\Phi}}{T_{i}}\sim\frac{k^{2}_{z}v_{Ti}^{2}}{\omega_{ci}^{2}}
\frac{1}{A^{1/2}_{in}\left(k_{\bot}^{2}\rho_{i}^{2}
\right)}\sim\frac{k_{z}\omega_{ci}}{k_{y}V'_{i0}}\frac{A^{1/2}_{in}\left(k_{\bot}^{2}\rho_{i}^{2}
\right)}{k_{\bot}\rho_{i}}.\label{67}
\end{eqnarray}
where Eq.(\ref{3}) was used. Level (\ref{67}) gives the estimate for
the amplitudes of the perturbed potential for different values of
the wave number $k$ for long wavelength IC waves. This level is
determined by balancing linear growth rate with nonlinear damping
rate, which we consider in our estimations as formed by
perturbations of the electrostatic potential with spectrum width
$\triangle\mathbf{k}\sim \mathbf{k}$ about $\mathbf{k}$. The
ultimate saturation level at which a steady state occurs over all
wave number space is determined as a level which is sufficient for
the stabilization of the wavenumber region of the modes with the
maximal growth rate. It follows from Eqs.(\ref{10}) that the growth
rate $\gamma_{02}\simeq\gamma_{e2}$ attains its maximal value at
$z_{2e}\sim 1$. Thus the ultimate saturation level is determined by
Eq.(\ref{67}) with $k_{z}/k_{\perp}$ determined by Eq.(\ref{52}) and
and it is equal to
\begin{eqnarray}
&\displaystyle\frac{e\widetilde{\Phi}}{T_{i}}\sim
\frac{v_{Ti}\omega_{ci}}{v_{Te}V'_{i0}}\frac{A^{1/2}_{in}\left(k_{\bot
0}^{2}\rho_{i}^{2} \right)}{k_{\bot 0}^{2}\rho_{i}^{2}},\label{68}
\end{eqnarray}
In Eq.(\ref{52}) $k_{\perp 0}\rho_{i}$ is determined by
Eq.(\ref{53}) and it is equal to
\begin{eqnarray}
&\displaystyle
k_{0\bot}\rho_{i}=\sqrt{2}\left(n!\sqrt{2}\frac{v_{Ti}V'_{i0}}{v_{Te}\omega_{ci}}\right)^{1/(2n-1)}.
\label{69}
\end{eqnarray}
Level (\ref{68}) is in maximum for the fundamental IC mode with
$n=1$, for which
\begin{eqnarray}
&\displaystyle
1>k_{0\bot}\rho_{i}\geq\frac{v_{Ti}V'_{i0}}{v_{Te}\omega_{ci}}
\label{70}
\end{eqnarray}
The condition $|z_{in}|>1$ of the weak ion cyclotron damping, under
which Eq.(\ref{42}) was obtained, imposes other restriction on the
admissible values for $k_{0\bot}\rho_{i}$, which for mode
$\delta\omega_{2}$ is
\begin{eqnarray}
&\displaystyle 1>k_{0\bot}\rho_{i}\geq\frac{\omega_{ci}}{V'_{i0}}.
\label{71}
\end{eqnarray}
The maximal level of IC turbulence, for which conditions (\ref{70})
and (\ref{71}) are met both, is of the order of
\begin{eqnarray}
&\displaystyle \frac{e\widetilde{\Phi}}{T_{i}}\sim
\left(\frac{\omega_{ci}}{V'_{i0}}\right)^{2} \label{72}
\end{eqnarray}
for
\begin{eqnarray}
&\displaystyle
\frac{\omega_{ci}}{V'_{i0}}<\left(\frac{v_{Ti}}{v_{Te}}\right)^{1/2},
\label{73}
\end{eqnarray}
and
\begin{eqnarray}
&\displaystyle \frac{e\widetilde{\Phi}}{T_{i}}\sim
\frac{v_{Ti}}{v_{Te}} \label{74}
\end{eqnarray}
when the condition opposite to (\ref{73}) met. Note, that levels
(\ref{72}) and (\ref{74}) appear to be much lower then level
(\ref{60}) for modified by shear flow current driven IC instability.

Now consider the short wavelength, $k_{\perp}\rho_{i}>1$, part of
the spectrum of IC waves. Accounting for Eqs.(\ref{52}) and
(\ref{53}) we obtain that
\begin{eqnarray}
&\displaystyle
\left|\frac{\gamma_{02}C_{2}}{C_{3}}\right|\sim\tau\left|\frac{k_{y}}{k_{z}}\frac{V'_{i0}}{\omega_{ci}}\right|
\frac{k^{4}_{z}\rho^{4}_{i}}{A_{in}\left(k^{2}_{\perp}\rho^{2}_{i}\right)}
\frac{k^{ 2}_{y}}{k^{2}_{z}}\gtrsim
\frac{v_{Ti}\omega_{ci}}{v_{Te}V'_{i0}}\gg 1. \label{75}
\end{eqnarray}
which furnishes the dominance of the $C_{2}$ terms in $\gamma_{sh}$
for that part of the spectrum. Nonlinear damping rate $\gamma_{sh}$
is equal now to
\begin{eqnarray}
&\displaystyle \gamma_{sh}\sim
\frac{6\gamma_{02}C_{2}}{\left(\delta\omega_{02}\right)^{2}}.
\label{76}
\end{eqnarray}
The estimation of the ratio $C_{1}/\gamma_{sh}$ with $\gamma_{sh}$
determined by (\ref{76}) gives,  that
\begin{eqnarray}
&\displaystyle\frac{C_{1}}{\gamma_{sh}}\sim
\frac{1}{k_{\perp}\rho_{i}}
\frac{k_{y}}{k_{z}}\frac{\omega_{ci}}{V'_{0i}}\sim\frac{v_{Te}}{v_{Ti}}\frac{\omega_{ci}}{V'_{0i}}\gg
1.\label{77}
\end{eqnarray}
Thus, the term $\gamma_{sh}$ in Eq.(\ref{49}) may be omitted. The
amplitude of the electrostatic potential $\widetilde{\Phi}$ in the
steady state may be determined in this case from the balance
equation $\gamma_{02}\left(\mathbf{k}\right)=C_{1}$, which yields
the estimate for the rms amplitudes of the perturbed potential for
different values of $k_{\perp}\rho_{i}$ for short wavelength IC
waves,
\begin{eqnarray}
&\displaystyle\frac{e\widetilde{\Phi}}{T_{i}}\sim\frac{v_{Ti}}{v_{Te}
}\frac{V'_{0i}}{\omega_{ci}}\frac{1}{\left(k_{\perp}\rho_{i}\right)^{1/2}},\label{78}
\end{eqnarray}
where the Eq.(\ref{52}) was used. Eq.(\ref{53}) and condition of
weak IC damping of IC waves impose on $k_{\perp}\rho_{i}$ the
restriction
\begin{eqnarray}
&\displaystyle \left(\frac{v_{Te}\omega_{ci}}{v_{Ti}V'_{0i}
}\right)^{1/2}\gtrsim k_{\perp}\rho_{i} \gtrsim
\frac{\omega_{ci}}{V'_{0i}}\label{79}
\end{eqnarray}
For that range of $k_{\perp}\rho_{i}$ level (\ref{78}) is in the
range
\begin{eqnarray}
&\displaystyle\frac{v_{Ti}}{v_{Te}}\left(\frac{V'_{0i}}{\omega_{ci}}\right)^{3/2}
\gtrsim\frac{e\widetilde{\Phi}}{T_{i}}\gtrsim
\left(\frac{v_{Ti}}{v_{Te}}\frac{V'_{0i}}{\omega_{ci}}\right)^{5/4},\label{80}
\end{eqnarray}
and the electric field strength
$\widetilde{E}_{\perp}=-k_{\perp}\widetilde{\Phi}$ is
\begin{eqnarray}
&\displaystyle \widetilde{E}_{\bot}\lesssim\frac{T_{i}}{e\rho_{i}}
\left(\frac{v_{Ti}}{v_{Te}}\frac{V'_{0i}}{\omega_{ci}}\right)^{3/4}.\label{81}
\end{eqnarray}
As a demonstration of level (\ref{81}) we take the data detected in
magnetopause by satellite Prognoz-8 \cite{Bleski,Belova},
$B=2\cdot10^{-3}$G, $V'_{0i}/\omega_{ci}=0.5$,  $T_{e}\sim
T_{i}=100$eV, and obtain the estimate $\widetilde{E}_{\bot}\lesssim
0.4\cdot10^{-3}$V/m at fundamental cyclotron mode, which is in good
agreement with measured value\cite{Belova}
$\widetilde{E}_{\bot}\simeq 0.5\cdot10^{-3}$V/m.

\subsection*{B.Strong flow shear. Shear flow driven ion cyclotron instabilities}
Under conditions of sufficiently strong flow shear, for which
condition opposite to (\ref{3}) is met, two shear flow driven IC
kinetic instabilities may be developed for
$k_{y}V'_{0}/k_{z}\omega_{ci}<0$ with frequency determined by
Eq.(\ref{11}). These instabilities are excited  with a growth rate
determined by (\ref{5}) and (\ref{6}) when
$k_{z}\left(V_{0e}-V_{0i}\right)>n\omega_{ci}$, and with growth rate
(\ref{9}) and (\ref{10}), when
$k_{z}\left(V_{0e}-V_{0i}\right)<n\omega_{ci}$. For short
wavelength, $k_{\perp}\rho_{i}>1$, ion cyclotron waves the estimate
\begin{eqnarray}
&\displaystyle\frac{\gamma_{01,02} C_{2}}{C_{3}}\sim \tau
\left(k_{\perp}\rho_{i}\right)^{1/2}
\left|\frac{k_{y}}{k_{z}}\frac{\omega_{ci}}{V'_{0i}}\right|^{1/2}>1\label{82}
\end{eqnarray}
is valid, which defines the dominance of the $C_{2}$ terms in
$\gamma_{sh}$ for that part of the spectrum. Nonlinear damping rate
$\gamma_{sh}$ in this case is equal to
\begin{eqnarray}
&\displaystyle\gamma_{sh}\simeq
-24\frac{\gamma_{01}C_{2}A_{in}\left(k_{\bot} ^{2} \rho _{i}^{2 }
\right)k_{z}^{2}v^{2}_{Ti}}{\left(1+\tau\right)\left(\delta\omega_{+}\right)^{4}}
\left(1+\left|\frac{k_{y}}{k_{z}}\frac{V'_{0i}}{\omega_{ci}}\right|\right).\label{83}
\end{eqnarray}
The estimation of the ratio $C_{1}/\gamma_{sh}$ with $\gamma_{sh}$
determined by (\ref{83}) gives  that
\begin{eqnarray}
&\displaystyle\frac{C_{1}}{\gamma_{sh}}\simeq
\frac{\left(1+\tau\right)}{24\tau}\left(k_{\perp}\rho_{i}\right)^{1/2}
\left(\frac{k_{y}}{k_{z}}\frac{\omega_{ci}}{V'_{0i}}\right)\gg
1.\label{84}
\end{eqnarray}
Therefore the term $\gamma_{sh}$ in Eq.(\ref{49}) may be omitted.
The amplitude of the electrostatic potential $\widetilde{\Phi}$ in
the steady state may be determined in this case from the balance
equation $\gamma_{0}\left(\mathbf{k}\right)=C_{1}$, which gives the
following estimate:
\begin{eqnarray}
&\displaystyle\frac{e\widetilde{\Phi}}{T_{i}}\sim\frac{1}{k_{\bot}
\rho_{i}}\left(\frac{k_{z}V'_{0i}}{k_{\perp}\omega_{ci}}\right)^{1/2}.\label{85}
\end{eqnarray}
By using Eqs.(\ref{52}) and (\ref{54}) we obtain from Eq.(\ref{85})
the ultimate estimate for the value of the perturbed potential in
the short wavelength part of the spectrum,
\begin{eqnarray}
&\displaystyle\frac{e\widetilde{\Phi}}{T_{i}}\sim\left(\frac{v_{Ti}V'_{0i}}{v_{Te}\omega_{ci}}\right)^{7/4}.\label{86}
\end{eqnarray}

Now we consider the long wavelength, $k_{\perp}\rho_{i}<1$, spectrum
subrange of IC waves. Comparing the contributions of $C_{2}$ and
$C_{3}$ to $\gamma_{sh}$ under condition of strong flow shear we
obtain that
\begin{eqnarray}
&\displaystyle\frac{\gamma_{01,02} C_{2}}{C_{3}}\sim \tau
\frac{k_{z}v_{Ti}}{\omega_{ci}}\frac{k^{2}_{y}}{k^{2}_{z}}
\left|\frac{k_{z}}{k_{y}}\frac{\omega_{ci}}{V'_{0i}}\right|^{1/2}\left(A_{in}\left(k_{\bot}
^{2}\rho_{i}^{2}\right)\right)^{1/2}\left(1-A_{i0}\left(k_{\bot}
^{2}\rho_{i}^{2}\right)+\tau+k^{2}\lambda^{2}_{Di}\right)^{1/2}\nonumber\\
&\displaystyle\lesssim \tau k^{2}_{\perp}\rho_{i}^{2
}\left(1-A_{i0}\left(k_{\bot}
^{2}\rho_{i}^{2}\right)+\tau+k^{2}\lambda^{2}_{Di}\right)^{1/2}.\label{87}
\end{eqnarray}
Thus for long wavelength, $k_{\perp}\rho_{i}<1$, ion cyclotron waves
the nonlinear growth rate $\gamma_{sh}$ in the case of strong flow
shear is dominated by the $C_{3}$ term and is approximated as
\begin{eqnarray}
&\displaystyle \gamma_{sh}\left(\mathbf{k}\right)\simeq
\frac{24}{1+\tau}C_{3}A_{in}\left(k_{\bot} ^{2} \rho _{i}^{2 }
\right)
\frac{k^{2}_{z}v^{2}_{Ti}}{\left(\delta\omega_{+}\right)^{4}}
\left|\frac{k_{y}V'_{0i}}{k_{z}\omega_{ci}}\right|,\label{88}
\end{eqnarray}
where the coefficient $C_{3}$ is determined from the equation
\begin{eqnarray}
&\displaystyle
C_{3}+\frac{2e^{2}}{m^{2}_{i}}\sum_{n_{1}=-\infty}^{\infty}\int
d\mathbf{k}_{1}|\Phi\left(\mathbf{k}_{1}\right)|^{2}
A_{in}\left(k_{1\bot}^{2}\rho_{i}^{2}
\right)k_{z}^{2}k_{1z}^{2}\left(\frac{k_{1y}V'_{i0}}
{k_{1z}\omega_{ci}}\right)^{2}
\frac{C_{3}}{\left(\delta\omega_{+}\right)^{4}}\nonumber\\
&\displaystyle=\frac{e^{2}}{3m^{2}_{i}}\sum_{n_{1}=-\infty}^{\infty}\int
d\mathbf{k}_{1}|\Phi\left(\mathbf{k}_{1}\right)|^{2}
A_{in}\left(k_{1\bot} ^{2} \rho _{i}^{2 }
\right)k_{z}^{2}k_{1z}^{2}\left(\frac{k_{1y}V'_{i0}}
{k_{1z}\omega_{ci}}\right)^{2}\frac{\gamma_{01}}{\left(\delta\omega_{+}\right)^{2}}.\label{89}
\end{eqnarray}
Remarkably, $\gamma_{sh}\left(\mathbf{k}\right)$ determined by
Eq.(\ref{88}) is positive and it may be responsible for the
development of the nonlinear instability. From the estimate of the
ratio $|C_{1}/\gamma_{sh}|$ under condition of strong flow shear
(\ref{54}),
\begin{eqnarray}
&\displaystyle \left|\frac{C_{1}}{\gamma_{sh}}\right|\sim
A_{in}\left(k_{\bot} ^{2} \rho _{i}^{2 }
\right)\frac{\tau^{2}\left(1+\tau\right)}{24}\left|\frac{k_{z}\omega_{ci}}{k_{y}V'_{0i}}\right|
<\frac{\tau^{2}\left(1+\tau\right)}{24}\frac{k^{2}_{z}v^{2}_{Ti}}
{n^{2}\omega^{2}_{ci}}\ll 1, \label{90}
\end{eqnarray}
it follows, that the dominant nonlinear term in Eq.(\ref{49}) in the
long wavelength part of the spectrum of IC waves in the case of
strong flow shear is $\gamma_{sh}$, determined by Eq.(\ref{88}) with
coefficient $C_{3}$, determined by Eq.(\ref{89}). Thus the combined
effect of strong shear flow and turbulent scattering of ions,
determined by the $\gamma_{sh}$, leads to the nonlinear instability
with growth rate (\ref{88}). Note, that the first term of the
left-hand side of Eq.(\ref{89}) dominates over the second when
\begin{eqnarray}
&\displaystyle\frac{e\widetilde{\Phi}}{T_{i}}\lesssim \tau
A_{in}^{1/2}\label{91}
\end{eqnarray}
and it becomes less than the second term at the level of turbulence
above (\ref{91}). At the level
\begin{eqnarray}
&\displaystyle\frac{e\widetilde{\Phi}}{T_{i}}\gtrsim \tau^{3/2}
A_{in}^{1/2}\label{92}
\end{eqnarray}
the nonlinear growth rate becomes greater than the linear growth
rate $\gamma_{0}\left(\mathbf{k}\right)$.

\section*{IV.DISCUSSION AND CONCLUSION.}

In this study, we have presented the renormalized theory of the IC
turbulence in the magnetic field--aligned plasma shear flow. The
developed theory extends the earlier studies \cite{Dum, Benford} of
the renormalized theory of the IC turbulence by including a new
combined effect of plasma turbulence and shear flow, determined in
nonlinear dispersion equation (\ref{43}) by the term
$\varepsilon_{sh}\left(\mathbf{k},\omega\right)$. This effect
consists in turbulent scattering of ions by the IC turbulence across
the shear flow into the regions with a greater or smaller flow
velocity and is the manifestation of the anomalous viscosity due to
ion cyclotron turbulence. Analytically it manifests in nonlinear
broadening of IC resonances. We have derived the approximate
solution (\ref{49}) of that equation and use it for qualitative
analysis of the role of the discovered effect in nonlinear evolution
of the shear flow modified current driven IC instability and shear
flow driven kinetic IC instabilities. We have considered the limits
of weak and strong flow shear for short wavelength and long
wavelength parts of the spectrum of IC waves separately and arrived
at the following conclusions.

1.The shear flow modified current driven IC instability, which
develops under condition of weak flow shear (\ref{3}), saturates as
the ordinary IC instability driven by shearless current on the high
level (\ref{60}) in the long wavelength, $k_{\bot}\rho_{i}\ll 1$,
part of the spectrum and on a very low level (\ref{58}) in the short
wavelength, $k_{\bot}\rho_{i}\gg 1$, part. The effect of the shear
flow, determined by the term $\gamma_{sh}\left(\mathbf{k}\right)$,
is negligible on the saturation of this instability.

2.The long wavelength part of the spectrum of the shear flow driven
IC instability under condition of weak flow shear saturates due to
combined effect of shear flow and turbulent scattering of ions,
determined by the term $\gamma_{sh}\left(\mathbf{k}\right)$ in
Eq.(\ref{48}), on levels (\ref{72}) or (\ref{74}) (depending on the
condition (\ref{73})), which appear to be much lower then the
corresponding level (\ref{60}) for current driven IC instability
modified by shear flow.

3.The saturation of the short wavelength spectrum subrange of the
shear flow driven instability under conditions of weak (\ref{49}),
as well as strong (\ref{49}) flow shear arises from the turbulent
scattering of ions by IC turbulence as in shearless plasma and
determined by term $C_{1}$ in Eq.(\ref{49}). It occurs at very low
levels (\ref{78}) and (\ref{86}), respectively, which are comparable
with corresponding level (\ref{58}) for shear flow modified IC
current driven instability.

4.Nonlinear evolution of the long wavelength part of the IC
turbulence spectrum developed by the shear flow driven instability
under conditions of strong flow shear is determined by the
$\gamma_{sh}\left(\mathbf{k}\right)$ term in Eq.(\ref{49}). In this
case combined effect of shear flow and turbulent scattering of ions
introduces a principally new effect into the nonlinear development
of the IC turbulence, which is absent in shearless plasma flows. It
is the \textit{shear flow driven nonlinear instability} with growth
rate (\ref{88}), which above level (\ref{91}) becomes greater than
the corresponding linear growth rate
$\gamma_{02}\left(\mathbf{k}\right)$, determined by Eqs.(\ref{9})
and (\ref{10}). This effect resembles the effect of the negative
viscosity. It contrasts dramatically with effect of ion scattering
by IC turbulence in shearless plasma flows\cite{Dum, Benford}, where
it leads only to saturation of IC instability.

It has to be noted here that we consider the model of the
collisionless plasma. However even weak ion-neutral collisions,
roughly $2\cdot10^{-2}$ of the IC frequency, are stabilizing. Ion
collision frequency may be absorbed into the linear growth rate,
leading to reducing the maximal growth rate. When ion-ion or
ion-neutral collision frequencies are even marginally greater than
an ion cyclotron instability growth rate, such instability does not
develop \cite{Satyanarayana}. The effect of ion collisions on the
evolution of the considered ion cyclotron instabilities in
near-Earth space plasmas determines the threshold altitude for the
instabilities development. Our collision-free analysis is completely
applicable to shear flows formed by solar wind around the Earth
magnetosphere, to auroral acceleration region where at an altitude
of 4000km the NASA's FAST (Fast Auroral SnapshoT) satellite observed
quite narrow ion beams\cite{McFadden} with $V'_{0}/\omega_{ci}$
ratio of the order of\cite{Amatucci-1999} $0.3$ for hydrogen ions
and of the order of $10$ for oxygen ions. The collision-free
analysis of our paper is pertinent to the F-region of the ionosphere
for altitudes above 180km. In this region ion-neutral collision
frequency $\nu_{in}$ is roughly $4\cdot10^{-3}$ of the ion cyclotron
frequency and this value diminishes with altitude. That value may
become less than the linear growth rate $\gamma_{e2}$
(Eq.(\ref{10})), $\gamma_{e2}\sim
\left(v_{Ti}/v_{Te}\right)^{2}\left(V'_{0}/\omega_{ci}\right)^{2}\left(k_{\bot}
^{2} \rho _{i}^{2 }/A_{in}\left(k_{\bot} ^{2} \rho _{i}^{2
}\right)\right)\omega_{ci}$, and is much less then
$\gamma_{\pm}\sim\left(v_{Ti}/v_{Te}\right)^{1/2}\left(V'_{0}/\omega_{ci}\right)^{1/2}\left(k_{\bot}
\rho_{i}A_{in}\left(k_{\bot}^{2}\rho_{i}^{2
}\right)\right)^{1/2}\omega_{ci}$ (see Eqs.(\ref{10}) and
(\ref{11})) for $\tau\sim 1$, $k_{\bot}\rho_{i}\sim 1$ for shear
flows with $V'_{0}\gtrsim\omega_{ci}$.

This paper is the first attempt of the analysis of the nonlinear
evolution of the IC turbulence of magnetic field--aligned  plasma
shear flow. Here we have considered only the effect of scattering of
ions by the IC turbulence in shear flow on the nonlinear evolution
of kinetic IC instabilities. It is well known that other nonlinear
effects, such as quasilinear flattening of the electron distribution
function in wave-particle resonance region\cite{Drummond-1962}, and
weak turbulence effects\cite {VS-KN1981} may change the ultimate
picture of the nonlinear stage of the IC turbulence. Real process of
the nonlinear evolution of the instabilities considered will include
the interplay of all these processes. That defines the need in the
development of additional nonlinear theories for the plasma shear
flows, which, however, are absent today. These are IC weak
turbulence theory for plasma shear flow, which has to include the
processes of the induced scattering of IC waves on ions and
electrons and processes of the decay of IC waves, ordinary and
renormalized quasilinear theories for ion and electron distribution
functions. Adequate coverage of such a broad area of nonlinear
plasma physics is difficult in a limited-page article and much more
have to be done for the development of a comprehensive nonlinear
theory of the turbulent state of plasma shear flows.
\section*{ACKNOWLEDGEMENTS}
One of the authors (VSM) gratefully acknowledge useful discussions
with Profs. R.Z.Sagdeev and P.H.Diamond. This work was supported by
Fundamental Researches State Fund of Ukraine under Project No
25.2/153.

\end{document}